\documentclass[aps,final,notitlepage,oneside,twocolumn,nobibnotes,nofootinbib,
superscriptaddress,noshowpacs,centertags]{revtex4-1}

\usepackage[utf8]{inputenc}
\usepackage[english]{babel}
\usepackage{graphicx}
\usepackage{latexsym}
\usepackage{amssymb}
\usepackage{amsmath}
\usepackage{float}

\begin{document}

\title{Spin of primordial black holes in the model with collapsing domain walls}
\author{Yury N. Eroshenko}\thanks{e-mail: eroshenko@inr.ac.ru}
\affiliation{Institute for Nuclear Research, Russian Academy of Sciences, pr. 60-letiya Oktyabrya 7a, Moscow, 117312 Russia}

\date{\today}

\begin{abstract}
The angular momentum (spin) acquisition by a collapsing domain wall at the cosmological radiation-dominated stage is investigated. During the collapses, primordial black holes and their clusters can be born in various mass ranges.  Spin accumulation occurs under the influence of tidal gravitational perturbations from the surrounding density inhomogeneities at the epoch when the domain wall crosses the cosmological horizon. It is shown that the dimensionless spin parameter can have the small values $a_S<1$ only for primordial black holes with masses $M>10^{-3}M_\odot$, whereas less massive black holes receive extreme spins $a_S\simeq1$. It is possible that primordial black holes obtain an additional spin due to the vector mode of perturbations.
\end{abstract}

\maketitle 





\section{Introduction}

Observations of gravitational wave bursts by LIGO/Virgo detectors revealed two interesting features of the merging objects population: relatively large masses and small angular momentum (spin\footnote{Although  it is more correctly to call by the ``spin'' an intrinsic angular momentum of elementary particles, according to the terminology accepted in astrophysics, the angular momentum of black holes is also named spin. This is justified by the fact that the external metric of a stationary black hole is characterized only by mass, angular momentum and charge, regardless of their origin.}) of individual objects \cite{FerPro19}. In many cases, the measured masses significantly exceed the typical masses of  compact stellar remnants. This can easily be explained by the selection effect, when the mergers of the most massive objects are more accessible for observations. This was predicted even before the registration of LIGO/Virgo events in the population synthesis \cite{Grietal01} and for massive stars in  the low metallicity regions \cite{chris1}. 
The measured effective spin turned out to be significantly less than the spin that occurs during the collapses of the star cores \cite{FerPro19, GarSilMor21,Beletal20}. However, the small spin can also be explained in the processes of ordinary stellar evolution \cite{Beletal20,PosKurMit19}.

Nevertheless, as noted in many papers (see e.g. \cite{FerPro19}), these features may also indicate the primordial origin of LIGO/Virgo black holes. In the most popular mechanisms the primordial black holes (PBHs) burn with small spins, and their masses can be distributed over a wide range. In addition, the primordial origin of the LIGO/Virgo  black holes is indicated by their mass function, which is close in shape to the log-normal  function \cite{Posetal20}.  

The possibility of the  PBH birth in the early Universe was predicted by Ya.B.~Zeldovich and I.D.~Novikov \cite{ZelNov67} and later it was considered by S.~Hawking \cite{Haw71}. To date, many possible mechanisms of their formation have been proposed \cite{KhlPol0,BerKuzTka83,DolSil93,RubKhlSak00,RubSakKhl01,Pietal18}, see, for example, the review \cite{CarKuh20}.  In the model of adiabatic (curvature) perturbation collapse the spin of the resulting PBHs has been calculated in a number of papers 
\cite{ChiYok17,MirGruNor19,Lucetal19,Haretal20}. But in the model of collapsing domain walls \cite{RubKhlSak00,RubSakKhl01,DenVil17,LiuGuoCai20}  calculation of the spin, to the best of our knowledge, has not been carried out before, and this work is devoted to such a calculation. 

In the model of  adiabatic perturbation collapses at the radiation-dominated stage, the PBH spin  is small, primarily because there is a strong correlation between the quadrupole moment of the collapsing region and the external mass distribution creating the tidal forces, and the high density peaks tends to be spherically symmetric. On the contrary, large spins are expected during the PBH formation at the early dust stages \cite{Haretal17,Kuh20}, but such PBHs should have small masses. We will show that during the collapses of the domain walls, a PBH can have a spin of various sizes, depending on the mass of the PBH and the parameters of the model. 

Measuring the masses and spins of black holes can help in determining whether they were born in astrophysical processes or have some kind of primordial origin. It should also be noted that theoretically, in addition to the primordial ones, there could also be ``relict'' black holes left over from previous epochs in the cyclic  Universe, if the Universe follows such a scenario \cite{GorTul21}. The spins and masses of the relict black holes could vary greatly due to accretion of the surrounding matter at the stage of the Universe compression.

A model of  the PBH formation from domain walls was proposed in \cite{RubKhlSak00,RubSakKhl01}. In addition to individual PBHs, clusters of PBHs with a fractal structure can also be born by this mechanism. However, in this paper we consider only isolated PBHs born outside clusters, or central PBHs of clusters dominating by mass.  The mass function of the PBH formed as a result of the domain wall collapses  was calculated in \cite{KirRub21}.

It is possible to distinguish three sources of angular momentum of a PBH: 1. vortex configurations of the scalar field arising at the stage of inflation; 2. tidal interaction of the quadrupole moment with the environment at the stage of collapse; 3. if the PBHs are formed in clusters, then the spin can change during the mergers of the PBHs, because the orbital angular momentum is partially converted into spin of the black holes formed during mergers \cite{JarGar21}.  In this paper, only the 2nd of these mechanisms is considered and the 1st is briefly discussed.

The angular momentum of galaxies arising under the influence of external tidal forces has been calculated in many works, \cite{Pee69}, 
\cite{Dor70}, etc. A useful approximation is the model of a homogeneous  (constant volume density) ellipsoid. In particular, this model was used in \cite{EisLoe95} to calculate the angular momentum of a gas cloud collapsing into a black hole. This mechanism is supposed to be the way of quasar black holes formation. In \cite{Lucetal19}, the approximation by ellipsoid was used in the model of the PBH formation from adiabatic perturbations. We also use an ellipsoidal approximation, but we are not considering a homogeneous ellipsoid, but a domain wall --- the surface of an ellipsoid with a constant surface density. The tidal forces twisting the wall are created by the density perturbations in the radiation, which scatter like sound waves after horizon crossing, so the spin is accumulated approximately during one Hubble time after the wall crosses the cosmological horizon. 

This article is organized as follows. The Section \ref{formsec} briefly describes the process of  the domain wall formation and their collapse into the PBHs. In the Section \ref{tidmsec}, the spin of the PBH obtained under the influence of tidal interactions with the environment is calculated. The Section \ref{primsec} discusses the possible role of the vector mode of perturbations in the appearance of the spin. Finally, the Section \ref{finsec} gives a discussion of the results obtained.


\section{Formation of domain walls and PBHs}
\label{formsec}

The evolution of closed domain walls and their collapse into black holes has been considered in a number of papers, starting from \cite{BerKuzTka83}. In the works \cite{RubKhlSak00,RubSakKhl01}, a specific mechanism for the formation of such walls at the inflation stage was proposed, this mechanism is describes in details in the monograph \cite{KhlRub04}. A complex scalar field  was considered with the Lagrangian
\begin{equation}
\mathfrak{L}=\frac{1}{2}\,\partial^\mu\phi\partial_\mu\phi^*-V,
\label{lagr}
\end{equation}
where the potential
\begin{equation}
V=\lambda\left(\phi\phi^*-\frac{f^2}{2}\right)^2+\Lambda^4(1-\cos\theta).
\label{veq}
\end{equation}
In this expression the scalar field is expressed in the form $\phi=f\exp{i\theta}/\sqrt{2}$, where 
$\theta(\vec x)=a_G(\vec x)/f$, and $a_G(\vec x)$ is the angular Goldstone mode, e.g axion. The second term in (\ref{veq}) results from the instanton effects which remove the vacuum degeneracy, $\Lambda$ is the parameter which in some models (e.g. for standard axions) may corresponds to the QCD energy scale. But in the model \cite{RubKhlSak00,RubSakKhl01} the parameter $\Lambda$ remains free and can be fixed by observational data if some populations of astrophysical black holes are the PBHs formed by this mechanism. The parameter $f$ is near the scale of the $U(1)$ symmetry breaking, and \cite{RubKhlSak00,RubSakKhl01} consider $f=1.77H$, where $H\sim10^{13}$~GeV is the Hubble parameter at the inflation stage.  

At the stage of exponential inflating in one e-fold, the difference of $\theta$ parameters on one Hubble scale is $\delta\theta=H/(2\pi f)$, and over time in some regions, the value of $\theta$ from the minimum of the potential $\theta=0$, stochastically evolving, can pass through the values of $\theta=\pi$ and begin to oscillate near the next minimum $\theta=2\pi$. Thus, rare regions with $\theta\simeq2\pi$ appear, surrounded by the regions with $\theta\simeq0$. In this case, there must be a closed boundary surface with $\theta=\pi$, which is a domain wall carrying a surface energy density
\begin{equation}
\sigma=4\Lambda^2f.
\label{sigeq}
\end{equation}
For a detailed description of these processes, see \cite{RubKhlSak00,RubSakKhl01,KirRub21,KhlRub04}.

The domain wall has a significant surface tension, but its compression begins only after the wall crosses the cosmological horizon, when the radius of the wall is $R\sim c/H=2ct$ (at the radiation-dominated stage). At this moment, the total mass of the wall is fixed $M=4\pi R^2\sigma$.  The process of the wall evolution is significantly influenced by its interaction with plasma \cite{RubSakKhl01,KurRub20}. Gravitational collapse into a black hole is possible when the wall thickness is less than its gravitational radius. Prior to the horizon crossing, the wall consists of causally unrelated parts, which themselves could collapse into smaller-mass PBHs with the formation of a  PBH cluster. In this paper, however, we do not consider PBH clusters. 

The surface density (\ref{sigeq}) can vary quite widely. PBH in astrophysically interesting mass intervals $10^{25}-10^{40}$~g, according to the work \cite{RubSakKhl01}, can be formed at $\Lambda\sim1-5$~GeV$^3$. This corresponds to $\sigma\sim (7\times10^{13}-2\times10^{15})$~GeV$^3$.


\section{Spin accumulation under the influence of tidal forces}
\label{tidmsec}

The domain wall is non-spherical, it has a quadrupole moment. Gravitational forces from the surrounding adiabatic perturbations create a moment of forces that causes the PBH spin. The collapsing domain wall is an entropic perturbation on the background of ordinary adiabatic perturbations, and both types of perturbations are mutually statistically independent to high degree. Really, adiabatic (curvature) perturbations originate from quantum fluctuations of some scalar field, inflaton. But the scalar field leading to the walls may be a spectator field. In this case (which is assumed in this work) the fluctuations of two different fields are statistically independent up to higher orders, and there should be no strong correlations between the directions of the main axes of the wall quadrupole moment and the distribution of the surrounding matter. It is also important that the spin accumulation occurs mainly at the epoch when the wall crosses the cosmological horizon, because at an earlier time the wall and its surroundings are causally disconnected, and at a later time the external tidal forces and the radius of the wall decrease rapidly. Therefore, spin is effectively gained in about one Hubble time around the moment of the horizon crossing. 

The collapsing wall is modelled by the surface of an ellipsoid with a constant surface density $\sigma$:
\begin{equation}
\frac{r_x^2}{a^2}+\frac{r_y^2}{b^2}+\frac{r_z^2}{c^2}=1.
\end{equation}
Normal to the surface is $\vec n=(2r_x/a^2,2r_y/b^2,2r_z/c^2)$, and the surface area element at the radius $\vec r=(r_x,r_y,r_z)$ is
\begin{equation}
dS=r^2d\Omega\frac{nr}{(\vec n\cdot\vec r)},
\end{equation}
where $d\Omega=d\phi\sin\theta d\theta$, and the multiplier takes into account that the normal to the surface is not directed along the $\vec r$.   

The spin accumulation occurs after the wall has entered under the cosmological horizon. At the initial stage of compression, the wall is at rest, so it is a weakly relativistic object. The running of the surrounding perturbation as sound waves in radiation occurs at a speed $c/\sqrt{3}$. This means that the change in the configuration of external tidal forces occurs slowly than the expansion of the horizon, and therefore the times under consideration capture the epoch of Newtonian theory applicability. This makes it possible to do a reliable estimates using the equations of Newtonian gravity.
Tidal forces with potential $\phi_{sh}$ exert a torque on the wall
\begin{equation}
\vec K=-\int dS\,\sigma[\vec r\times\nabla\phi_{sh}].
\end{equation}
We neglect the rotation of the main axes of the ellipsoid during the angular momentum accumulation. The potential of the outer region $r>R$, which creates tidal forces, can be decomposed into spherical harmonics, as it was done in \cite{EisLoe95}:
\begin{equation}
\phi_{sh}=\sum_{l,m}\frac{4\pi G}{2l+1} a_{lm}Y_{lm}|\vec r|^l,
\label{tidpot}
\end{equation}
where
\begin{equation}
a_{lm}=-\bar\rho\int\limits_{|s|>R}d^3sY_{lm}^* \delta(\vec
s)s^{-l-1},\label{alm}
\end{equation}
$G$ --- gravitational constant,
$\rho(\vec r)$ --- density at the point $\vec r$, $\bar\rho$ --- average
cosmological density, $\delta(\vec r)\equiv (\rho(\vec r)-\bar\rho)/\bar\rho$.
The term with $l=1$ leads only to the movement of the center of mass of the wall and does not affect  the angular momentum, and the terms with $l\ge3$ can
also be neglected. Correlator
\begin{equation}
\langle a_{2m}a^*_{2n}\rangle=\frac{\bar\rho^2\sigma^2_{\rm H}(M)}{2\pi^2}\delta_{mn}.
\label{a2mkorr}
\end{equation}
At the moment of the horizon crossing, the r.m.s. perturbation at the mass scale $M$ is \cite{GreLid97}
\begin{equation}
\sigma_{\rm H}(M)\simeq9.5\times10^{-5}\left(\frac{M}{10^{56}
\mbox{~r}}\right)^{\frac{1-n_s}{4}},
\label{normsigmah}
\end{equation}
where $n_s$ is the power index of the spectrum of adiabatic perturbation obtained from CMB observations, $n_s=0.9608\pm0.0054$.

\begin{figure}[tbp]
\begin{center}
\includegraphics[angle=0,width=0.45\textwidth]{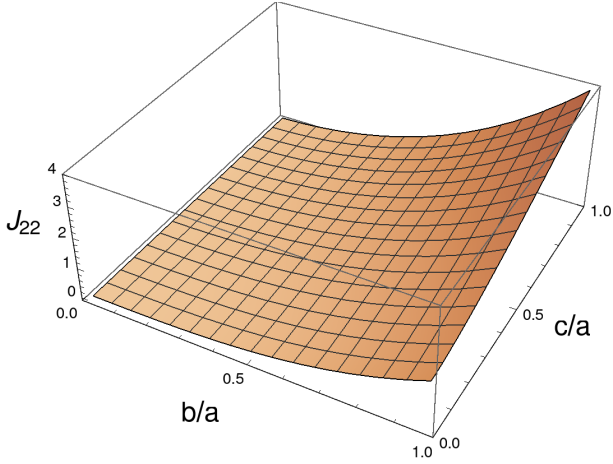}
\end{center}
\caption{Function $J_{22}$ depending on the semi-axes fractions $b/a$ and $c/a$.}\label{gr3d}
\end{figure}

In \cite{EisLoe95} (Eq.~(14)) the matrix $\Phi_{\alpha\beta}$ of the quadratic part $\phi_{sh}=(1/2)\Phi_{\alpha\beta}r_\alpha r_\beta$ of the tidal potential was also calculated. The $\alpha$ component of the torque can then be written as
\begin{equation}
K_\alpha=-\frac{1}{2}\sigma e_{\alpha\beta\gamma}\Phi_{\gamma\varkappa}\int r^3 d\Omega\left(\frac{r_x^2}{a^4}+\frac{r_y^2}{b^4}+\frac{r_z^2}{c^4}\right)^{1/2}r_\beta r_\varkappa
\end{equation}
(summation is performed by repeating indexes). Let 's rewrite it in the form
\begin{equation}
K_\alpha=-\frac{1}{2}\sigma e_{\alpha\beta\gamma}\Phi_{\gamma\varkappa}a^4J_{\beta\varkappa},
\end{equation}
where the matrix of dimensionless integrals $J_{\beta\varkappa}$ has a diagonal form, and we get its components by numerical integration at different ratios $b/a$ and $c/a$. See Fig.~\ref{gr3d}, where the function $J_{22}$ is shown for example.

Calculate the square
\begin{equation}
K_\alpha K_\alpha=\frac{1}{4}\sigma^2 a^8 {\rm Tr}\left\{J^2\Phi^2-(\Phi J)^2\right\}.
\label{trexpr}
\end{equation}
The dynamics of the domain wall compression is a separate complex problem \cite{KurRub20}. As an initial estimate, we assume that the compression occurs in one Hubble time $t_H$. Then the accumulated angular momentum
\begin{equation}
L\simeq (K_\alpha K_\alpha)^{1/2}t_H.
\end{equation}
When calculating matrix spur in (\ref{trexpr}), we use the form of the matrix $\Phi_{\alpha\beta}$ from the work \cite{EisLoe95} and get the following compact result for the dimensionless PBH spin
\begin{equation}
a_S=\frac{cL}{GM_{\rm PBH}^2}\simeq\frac{k_0c^4F\sigma_{\rm H}}{G^2\sigma M_H},
\end{equation}
where $k_0=2^{-19/2}3^{3/2}5^{1/2}\pi^{-7/2}\simeq2.9\times10^{-4}$, $M_H=t_Hc^3/G$ is the mass of the radiation inside the horizon at the moment $t_H$, and 
the notation 
\begin{equation}
F=[\alpha^2(J_{11}-J_{22})^2+\beta^2(J_{11}-J_{33})^2+\gamma^2(J_{22}-J_{33})^2]^{1/2}
\end{equation}
is used. The components $J_{\alpha\beta}$, in turn, depend on the relations $b/a$ and $c/a$, and the values $\alpha$, $\beta$, $\gamma$ have a Gaussian probability distribution arising from the Gaussian distribution of cosmological density perturbations,
\begin{equation}
P_1d\alpha d\beta d\gamma=\frac{1}{(2\pi)^{3/2}}e^{-\alpha^2/2-\beta^2/2-\gamma^2/2}d\alpha d\beta d\gamma.
\end{equation}

\begin{figure}[tbp]
\begin{center}
\includegraphics[angle=0,width=0.45\textwidth]{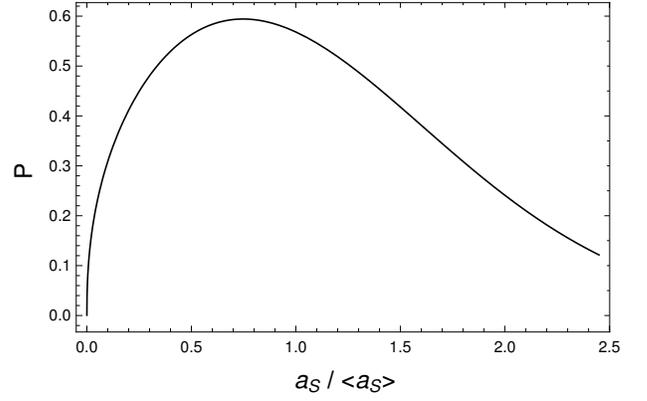}
\end{center}
\caption{Probability distribution of various values of the spin of the PBH normalized by the average spin.} \label{grprob}
\end{figure}

In the work \cite{RubSakKhl01} it is shown that the domain wall during its formation at the inflation stage randomly takes an arbitrary shape (we approximate it with the surface of an ellipsoid). Therefore, for a given large semi-axis of the ellipsoid, the probability distribution for the other axes can be approximated as a flat distribution
\begin{equation}
P_2d\left(\frac{b}{a}\right)=\theta\left(1-\frac{b}{a}\right)d\left(\frac{b}{a}\right),
\end{equation}
where $\theta(...)$ is a theta function, and similarly for $c/a$.

Averaging with $P_1$ and $P_2$ gives for $F$ the average value of $\bar F=1.05$. Therefore, the average spin value is
\begin{equation}
\langle a_S\rangle\simeq\frac{k_0c^4\bar F\sigma_{\rm H}}{G^2\sigma M_H}.
\label{itog}
\end{equation}
Numerically
\begin{eqnarray}
\langle a_S\rangle&\simeq& 10^{-4}\left(\frac{\bar F}{1.05}\right)\left(\frac{\sigma_{\rm H}}{6\times10^{-5}}\right)\left(\frac{M_{\rm PBH}}{30M_\odot}\right)^{-1}
\nonumber
\\
&\times&f_{\rm H}\left(\frac{\sigma}{10^{14}\mbox{~GeV$^3$}}\right)^{-1},
\label{itog2}
\end{eqnarray}
where$f_{\rm H}=M_{\rm PBH}/M_H$, and the numerical value $\sigma_{\rm H}=6\times10^{-5}$ corresponds to the mass inside the horizon $M_H=30M_\odot$ according to (\ref{normsigmah}). We suppose that the PBH mass is a fixed fraction of $M_H$, i.e. $f_H=const$. Ideally, the entire wall collapses if gravitational radiation is neglected, in this case $f_H\simeq1$.

We find the distribution of the spin values numerically by dividing the parameter space into discrete bins. The resulting distribution is shown in Fig.~\ref{grprob}.

As can be seen from (\ref{itog2}), the spin of the PBH does not reach the extreme value $a_S=1$ only for black holes with masses $M_{\rm PBH}>10^{-3}M_\odot$. For PBH of smaller masses, tidal forces twist the PBH to extreme spin values. Such an effective accumulation of the spin at low masses is explained by the fact that during the collapse of the domain walls, unlike the collapses of adiabatic perturbations, there is no suppression associated with the correlation of the quadrupole moment directions of internal and external matter. In addition, in the model of adiabatic perturbation collapses, PBHs are formed from the high density peaks $\nu=\delta/\sigma_{\rm H}\gg1$, the shape of which is close to spherically symmetric, as shown in the theory of Gaussian random fields. On the contrary, in our case, the shape of the domain walls is predominantly non-spherical, so they have large quadrupole moments. However, in the case of domain walls at large masses, the fact that the mass is concentrated on the surface, and not in the volume begins to play a role. Qualitatively, this can be explained as follows. In the case of curvature perturbation collapse, the time of the collapse is associated with the PBH mass by the unambiguous relation $t_H\simeq GM/c^3$, following from the condition $M\sim (4\pi/3)\rho_r(t)(2ct_H)^3$ at the horizon crossing moment $t_H$. However, for the domain walls $M\sim 4\pi\sigma (2ct_H)^2$, therefore the spin dependence on the mass is different, and (if other parameters are fixed), the spin is $\propto 1/M$. Therefore, starting from a certain mass, the spin becomes small. In the case of curvature perturbation, there would be no such dependence on mass. Indeed, if we formally replace $\sigma\to R\rho_r=2ct_H\rho_r(t_H)\propto 1/t_H$ with $t_H\simeq GM/c^3$, then the masses $M$ in the final expression will cancel. By this substitution we have switched roughly  to the curvature perturbation collapse model.

The functional form of the final result (\ref{itog2}) weakly depends on the domain walls formation model, under the conditions that the shape distribution $P_2d(b/a)$ is close to flat, and that the walls begin to collapse just after the horizon crossing. Indeed, the calculation used only the assumption that the domain walls have a constant surface density $\sigma$, and that the accumulation of the spin occurs near the the horizon crossing epoch. These assumptions are fulfilled for the model \cite{RubKhlSak00,RubSakKhl01}, however, if the domain walls are formed by some other mechanism, then they may be more spherically symmetric and have a smaller quadrupole moment. As for numerical results, even for the model \cite{RubKhlSak00,RubSakKhl01} there is a strong dependence on the particular values of the parameters of Lagrangian (\ref{lagr}). In \cite{RubKhlSak00,RubSakKhl01}, these parameters are fixed by the conditions that the PBH masses fall into the astrophysically interesting intervals, but formally these parameters can vary by many orders of magnitude.

We calculated the spin under the simplest assumptions about the shape of the perturbation spectrum. The shape depends on the properties of the inflationary potential and is a priori unknown. Since the considered scales are several orders of magnitude smaller than those from which galaxies are formed, the decrease or increase in perturbations on the small scales does not contradict the available observational data. Thus, we cannot rule out that the spins of low-mass PBH can have other values. It is possible also that the presence of spin to some extent delays the final collapse of the domain wall.

After the formation of PBH, its spin could change under the influence of scattered sound waves in radiation, as well as a result of accretion. However, in \cite{ChiYok17} it was shown that these two effects are weak and do not lead to a noticeable change in spin. However, in a recent paper \cite{Lucetal20}, it was shown that if the PBHs are heavier than around $10M_\odot$, accretion can become relevant at lower redshift potentially spinning up the PBHs.


\section{Vector mode}
\label{primsec}

As is known, during the inflation, three modes of perturbations can be generated: scalar, vector and tensor. The domain wall formed as a result of the phase transition can be twisted initially from the moment of its birth and carry a vector mode. We will call this vector mode the primordial spin. 
The relativistic theory of  mode evolution  in an ideal fluid is usually considered and it was concluded that the vector mode at the radiation-dominated stage decays rapidly, as $1/a^2(t)$. The vector mode does not play a role both for galaxies and for PBH formed by the collapses of adiabatic perturbations, since $a(t)$, as a rule, changes huge times from the end of inflation to collapse.

The attenuation of the vector mode was due to the fact that for an ideal fluid in the form of radiation, the condition $(\varepsilon +p)a^4=const$ was fulfilled. However, in our case, the carrier of the vector mode is not an ideal fluid, but a domain wall. Can the vector mode survive in this situation and contribute to the spin of a PBH? The main point is that the value of $4\pi\sigma R^2$ increases during the expansion of the Universe, because its radius is $R\propto a(t)$ increases until the horizon crossing. This property of the wall is due to its internal vacuum equation of state. And, if the carrier of the vector mode is the domain wall, then it is possible that the primordial spin is not lost and it gives a noticeable contribution to the PBH spin. However, to clarify this issue  the construction of a relativistic theory of the vector mode evolution carried by a domain wall is requires, which is beyond the scope of this work. 

To find the primordial spin  it is necessary to calculate the angular momentum tensor of the field configuration leading to the appearance of a domain wall at the inflation stage. In theory (\ref{lagr}), the calculation of the angular momentum tensor gives the spin value
\begin{equation}
\vec L=\frac{1}{2}\int [\vec x \times \nabla(\phi+\phi^*)]\,d^3x,
\end{equation}
where the integration covers the volume of the domain wall, but is effectively concentrated on its surface. In this case, the distribution of the scalar field should be found by solving a nontrivial problem of quantum fluctuations at the inflation stage for the field with the Lagrangian (\ref{lagr}).  
Next, it is required to calculate the evolution of the vector mode associated with the domain wall until its collapse into a PBH. This will allow one to find the contribution of the primordial spin to the total PBH angular momentum.

In this Section, we have only outlined some areas of primordial spin research. The corresponding calculations can be performed in future works.


\section{Conclusion}
\label{finsec}

In the model of collapsing adiabatic perturbations for dimensionless spin, the values $a_S<0.4$ \cite{ChiYok17}, $a_S\sim0.01$ \cite{MirGruNor19,Lucetal19}, and $a_S\sim O(10^{-3})$ \cite{Haretal20} were obtained. Ref.~\cite{ChiYok17} assumed a scaling relation for angular momentum to compute the
resulting spin after collapse of perturbations, while Refs.~\cite{MirGruNor19,Lucetal19,Haretal20} accurately computed the spin distribution from the properties
of perturbations in the early universe. In this paper, the PBH spin was calculated in the model of collapsing domain wall. The source of angular momentum is the external tidal gravitational forces interacting with the quadrupole moment of the wall. For the accumulation of angular momentum, it is important that the wall has a non-spherical shape, which is expected in the model of the wall birth  from inflationary vacuum fluctuations. It was obtained that the value of the dimensionless spin $a_S\simeq1$ for a PBH with masses $M\leq10^{-3}M_\odot$ (provided that the wall constitutes the fraction $f_{\rm H}\sim1$ of the total mass inside the horizon in the collapse epoch). 

The small value of the spin $a_S\ll 1$ obtained by the stellar mass PBH means that the LIGO/Virgo events with a small spin can also be explained by PBH generated as a result of domain wall collapses. For low-mass PBH in the early universe, spin may play an important role \cite{PacSil20}. For example, the spin of small-mass PBHs can be very important for the effect of their Hawking evaporation, since it affects the power and spectrum of radiation \cite{ArbAufSil19,ArbAufSil20,DasLahRay20}.
As for supermassive PBHs, they can serve as seeds of quasar black holes and significantly increase their masses due to disk accretion. As a result, the final black hole can carry a much larger dimensionless spin than the seed spin. The spin of the black hole at the center of the galaxy M87 is $a_S=0.75\pm0.15$ \cite{DokNaz19}, but this black hole could have increased its mass by many orders of magnitude due to gas accretion or mergers with other black holes. Therefore, the primordial origin of black holes in the centers of galaxies or their seeds does not meet contradictions in the magnitude of their spin.

It is not yet known for certain whether there are PBHs in the Universe. But even upper limits on their number in a particular mass range imposes useful restrictions on the spectrum of primordial perturbations and other effects \cite{CarKuh20}. If PBHs exist, they may be responsible for a number of processes in astrophysics and cosmology. In particular, as already mentioned, PBHs can explain some LIGO/Virgo events (except for events with neutron stars). In particular, the PBHs provide an explanation for events with small spin. Massive PBHs could provide seeds for the formation of quasars at redshifts $z>6$, while the ordinary astrophysical processes encounter difficulties in explaining the appearance of supermassive black holes in such early epochs. There are a number of other processes in which PBHs could play a significant role, for example, they could influence the absorption of relic radiation at 21~cm in the era of the ``Dark Ages'' \cite{Yan21}.

\acknowledgments

The author is grateful to S.G.~Rubin and A.A.~Kirillov for numerous useful discussions and to anonymous Referee for helpful comments.


\end{document}